Mn$_2$VAl Heusler alloy thin films: Appearance of antiferromagnetism and an exchange bias in a layered structure with Fe

Tomoki Tsuchiya[1,a)], Ryota Kobayashi[2], Takahide Kubota[3,4,b)], Kotaro Saito[5], Kanta Ono[5], Takashi Ohhara[6], Akiko Nakao[7], Koki Takanashi[3,4].

[1] Department of Material Science, Graduate School of Engineering, Tohoku University, Sendai 980-8579, Japan

[2] Department of Physics, Graduate School of Science, Tohoku University, Sendai 980-8578, Japan

[3] Institute for Materials Research, Tohoku University, Sendai 980-8577, Japan

[4] Center for Spintronics Research Network, Tohoku University, Sendai 980-8577, Japan

[5] High Energy Accelerator Research Organization (KEK), 1-1 Oho, Tsukuba 305-0801, Japan

[6] J-PARC Center, Japan Atomic Energy Agency, Tokai 319-1195, Japan,

[7] Research Center for Neutron Science and Technology, Comprehensive Research Organization for Science and Society, Tokai 319-1106, Japan

---

a) Author to whom correspondence should be addressed. Electronic mail: t.tomoki@imr.tohoku.ac.jp
b) Author to whom correspondence should be addressed. Electronic mail: tkubota@imr.tohoku.ac.jp

**(Abstract)**

Mn$_2$VAl Heusler alloy films were epitaxially grown on MgO(100) single crystal substrates by means of ultra-high-vacuum magnetron sputtering. A2 and L2$_1$ type Mn$_2$VAl order was controlled by the deposition temperatures. A2-type Mn$_2$VAl films showed no spontaneous magnetization and L2$_1$-type Mn$_2$VAl films showed ferrimagnetic behavior with a maximum saturation magnetization of 220 emu/cm$^3$ at room temperature. An antiferromagnetic reflection was observed with neutron diffraction at room temperature for an A2-type Mn$_2$VAl film deposited at 400ºC. A bilayer sample of the antiferromagnetic A2-type Mn$_2$VAl and Fe showed an exchange bias of 120 Oe at 10 K.

# I. INTRODUCTION

Exchange bias (EB) is the shift of a hysteresis loop caused by the interaction between an antiferromagnet and a ferromagnet, which was discovered by Meiklejohn and Bean in 1956 [1]. Since B. Dieny *et al*. applied EB for a spin valve structure [2], EB has been utilized in electronics such as magnetoresistive devices. To date, many researchers have focused on Ir-contained antiferromagnetic alloys which exhibit a large EB shift with high thermal stability [3-8]. However, the scarcity of iridium, which is the rarest element in the earth's crust [9], will be a critical issue for the stable supply of devices containing iridium in near future. Therefore, it is needed to find thermally stable antiferromagnets without iridium and also other scarce elements.

We focus on $Mn_2VAl$ Heusler alloy for the replacement of Ir-containing antiferromagnets. $L2_1$-type $Mn_2VAl$ is known as a ferrimagnetic full Heusler alloy [10-15]. Recently, however, magnetization and neutron diffraction measurements of a bulk sample have demonstrated that A2-type $Mn_2VAl$ is an antiferromagnet with a Néel temperature above 600 K [16]. There is an advantage for Heusler-type antiferromagnets from a view point of the lattice matching with ferromagnetic materials for spintronics, such as Co-Fe alloys, and cobalt-based Heusler alloys in which highly spin polarized transports are realized for the layered film samples with a (001)-texture [17-20]. In this paper, we present the antiferromagnetism in A2-type $Mn_2VAl$ thin films and an EB shift in a $Mn_2VAl$/Fe layered sample. The chemical phases of $Mn_2VAl$ films were successfully controlled by the deposition temperatures, and the magnetic properties were investigated by using laboratory facilities and neutron diffraction technique.

## II. EXPERIMENTAL PROCEDURES

All samples were deposited on 0.5 mm-thick MgO(100) single crystal substrates by using ultra-high-vacuum (UHV) magnetron sputtering technique. Surfaces of MgO substrates were cleaned by flushing at 700ºC in the UHV-chamber before sputtering. Subsequently, a $Mn_2VAl$ layer was deposited by using an alloy target with the composition of $Mn_{51}V_{25}Al_{24}$ (at.%) determined by inductively coupled plasma optical emission spectroscopy technique. The deposition temperatures ($T_{sub}$) were set at room temperature (RT) and temperatures from 300ºC to 800ºC in 100ºC increments. The thicknesses of $Mn_2VAl$ were 100 nm for structure determination and magnetization measurements and 1 μm for neutron diffraction measurements. A Ta capping layer with a thickness of 3 nm was deposited on the $Mn_2VAl$ layer at room temperature. The crystalline structures of $Mn_2VAl$ films were characterized by x-ray diffraction (XRD) and neutron diffraction. $Mn_2VAl$/Fe bilayers with a stacking structure of Ta (3 nm)/Fe (3 nm)/$Mn_2VAl$ (100 nm)//MgO(100) substrate were also prepared for EB measurements. $T_{sub}$ for the $Mn_2VAl$ layer was RT or 400ºC and that for Fe and Ta layers was RT. The bilayer samples were annealed at 200ºC with a magnetic field of 10 kOe in a vacuum furnace to induce an exchange anisotropy. Magnetization curves (*M*–*H* curves) were measured at RT and 10 K by using a vibrating sample magnetometer and a superconducting quantum interference device, respectively. Magnetic field was applied parallel to the <100> direction of the MgO substrate which corresponds to the <110> direction of epitaxially grown $Mn_2VAl$ and Fe layers. Neutron diffraction experiments were performed at BL18 SENJU, installed at the Materials and Life Science Experimental Facility in the Japan Accelerator Research Complex (MLF/J-PARC) [21]. The measurements were made in two sample settings where the epitaxial axis was parallel and perpendicular to the incident neutron beam to observe reflections in the ($hhl$) plane and the ($hk0$) plane of Mn$_2$VAl, respectively.

## III. RESULTS AND DISCUSSION

### A. Structural properties

Out-of-plane XRD patterns of the Mn$_2$VAl films are shown in Figure 1(a). All samples show 004 fundamental reflection and the samples deposited at 500 and 600ºC show 002 superlattice reflections. The $\Phi$ scan results for $T_{sub}$ = 500 and 600ºC are shown in Figure 1(b) and 1(c), respectively. Both samples exhibit the (111) superlattice reflections with four-fold symmetry. These results indicate that the samples deposited at 500 and 600ºC are L2$_1$-type Mn$_2$VAl and the other samples are A2-type Mn$_2$VAl. Note that all samples show the fundamental (220) reflections with four-fold symmetry representing an epitaxial growth of the Mn$_2$VAl layer. Lattice parameters along $a$- and $c$-axes calculated from 004 and 022 reflections are plotted in Figure 1(d) as a function of $T_{sub}$. The value of a bulk sample [10] is also shown for comparison. The in-plane lattice parameter of Mn$_2$VAl is expanded probably due to the lattice mismatch of 1.2% with the MgO(100) substrate. For $T_{sub}$ = 700ºC and 800℃, both lattice parameters are larger than those of the bulk sample, which may be attributed to a change of the film composition: The compositions of the Mn$_2$VAl layer for $T_{sub}$ = 700 and 800ºC are Mn$_{35.5}$V$_{32.3}$Al$_{32.2}$ and Mn$_{21.9}$V$_{39.1}$Al$_{39.0}$, respectively, whereas the stoichiometry of the Mn$_2$VAl layer is maintained for $T_{sub}$ ≤ 600ºC. This indicates that a high $T_{sub}$ leads to Mn sublimation, and such a composition change was also reported in a previous paper by Meinert *et al.* [14]. Long-range order parameters of Mn$_2$VAl for B2 and L2$_1$ phases, $S_{B2}$ and $S_{L21}$, are

plotted in Figure 1(e). $S_{B2}$ and $S_{L21}$ are estimated using the following equations:

$$S_{B2} = \{[I(002)_{exp}/I(004)_{exp}]/[I(002)_{sim}/I(004)_{sim}]\}^{1/2}, \quad (1)$$

$$S_{L21} = \{[I(111)_{exp}/I(022)_{exp}]/[I(111)_{sim}/I(022)_{sim}]\}^{1/2}, \quad (2)$$

where $I(hkl)_{exp(sim)}$ represents the experimental (simulated) integrated intensity of a ($hkl$) reflection. For $T_{sub} \leq 400°C$, the samples are A2 phase. Both $S_{B2}$ and $S_{L21}$ are more than 0.6 for $T_{sub}$ = 500 and 600°C. For $T_{sub}$ over ≥ 700°C, the samples become another phase due to the Mn sublimation.

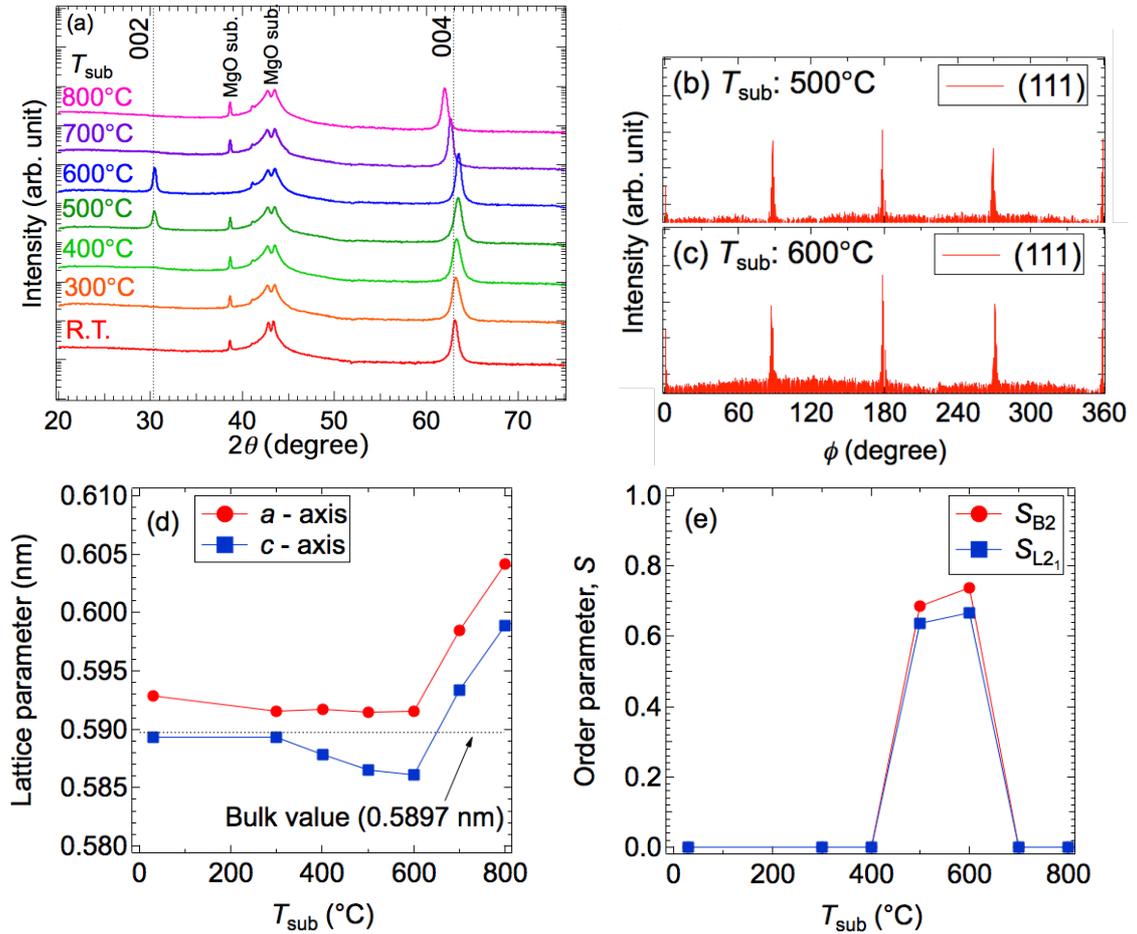

FIG. 1. Structural properties of 100 nm-thick $Mn_2VAl$ films. (a) Out-of-plane XRD

patterns, $\Phi$ scan results for (111) reflections of the Mn$_2$VAl layer deposited at (b) 500 and (c) 600°C, (d) the lattice parameters for in-plane (*a*-axis) and perpendicular-to-plane (*c*-axis) directions, and (d) order parameters for B2 phase ($S_{B2}$) and L2$_1$ phase ($S_{L21}$) as a function of the deposition temperature, $T_{sub}$.

## B. Magnetic properties of Mn$_2$VAl films

*M–H* curves of the Mn$_2$VAl films measured at RT are shown in Figure 2. The maximum applied magnetic field was 20 kOe. No hysteresis is observed for $T_{sub}$ less than 400°C and more than 700°C showing that A2-type Mn$_2$VAl is a paramagnet or an antiferromagnet. On the other hand, the samples with $T_{sub}$ = 500 and 600°C exhibit hysteresis. While the trends in the $T_{sub}$ dependence of the crystal structure and the saturation magnetization are similar to those reported in the previous paper [13], our samples show larger $S_{B2}$, $S_{L21}$, and $M_s$. This may be explained by the improved stoichiometry of the film composition.

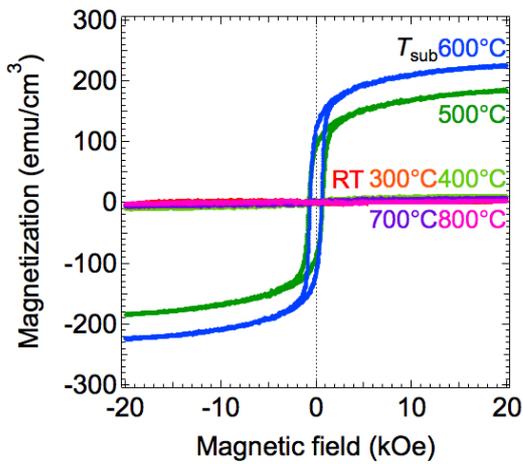

FIG. 2. *M-H* curves of 100 nm-thick Mn$_2$VAl films measured at RT

## C. Neutron diffraction experiments

In order to identify the magnetism of A2-type $Mn_2VAl$, neutron diffraction experiments at RT were performed using 1-μm-thick $Mn_2VAl$ films with $T_{sub}$ = RT, 400°C, and 600°C. Figures 3(a) – 3(c) show reflections in the (*hhl*) plane of the $Mn_2VAl$ films from which one can determine the chemical phases of the samples. Schematic illustrations of the expected reflections from A2 and $L2_1$ phases are shown in fig. 3(d), together with those from MgO. Fig 3(e) depicts the epitaxial relationship between $Mn_2VAl$ and MgO. Note that the broken lines in fig. 3(a)-3(c) and the solid lines in fig. 3(d) illustrate integer indices of MgO, while the broken lines in fig. 3(d) illustrate those of $L2_1$ phase. For $T_{sub}$ = 600°C, apart from strong spots from MgO, reflections with all-even and all-odd Miller indices with $L2_1$ notation were observed (blue arrows in fig. 3(a) and 3(d)), showing that the sample has $L2_1$ structure. For $T_{sub}$ = 400°C and RT, on the other hand, only reflections from the A2 phase (red arrows in fig. 3(b) and 3(c)) were confirmed, showing that these samples have the A2 structure.

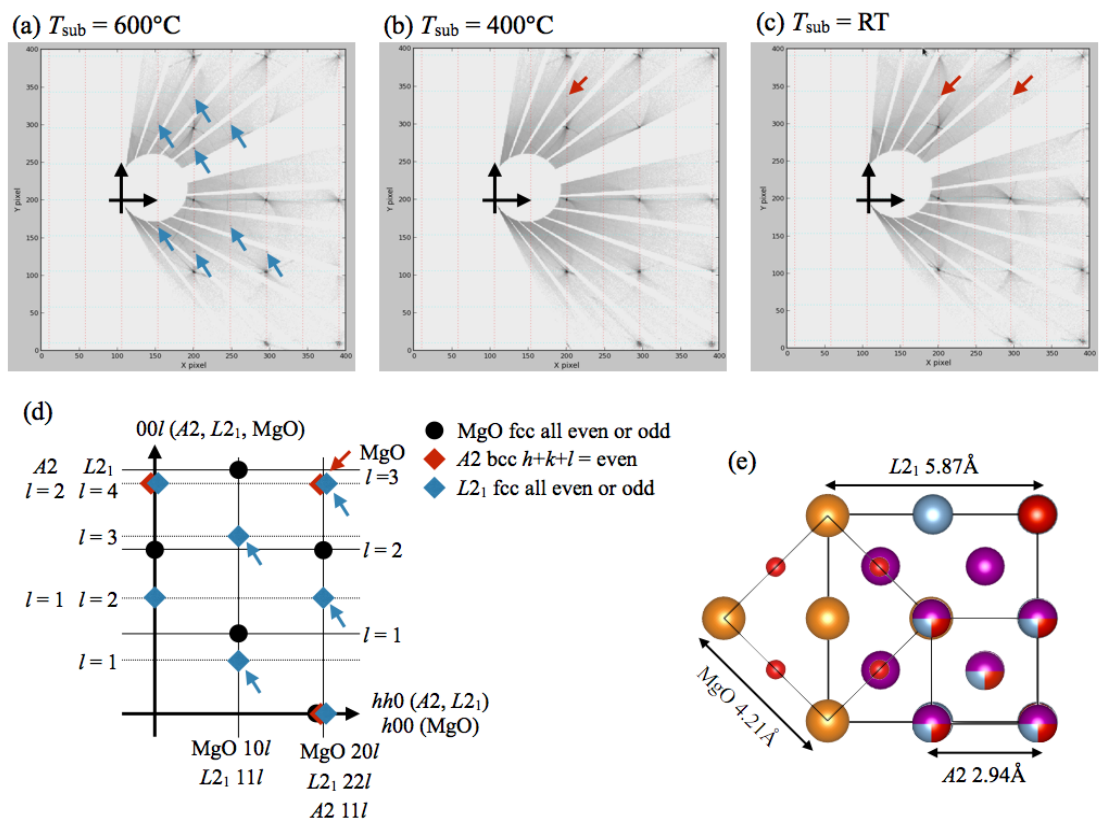

FIG. 3 Bragg reflections in the $(hhl)$ planes of A2 and L2$_1$ phases. (a)-(c) Observed reflections for Mn$_2$VAl films with $T_{sub}$ = 600°C, 400°C, and RT. (d) Schematic illustration of the observed reflections indicated by blue and red arrows for L2$_1$ and A2 phases, respectively. (e) Epitaxial relationship between Mn$_2$VAl and MgO.

Figures 4(a) – 4(c) show reflections in the $(hk0)$ plane of the Mn$_2$VAl films. Schematic illustrations of the expected reflections are shown in fig. 4(d) and 4(e) in which blue and red arrows indicate observed reflections. A reflection corresponding to 100 in A2 phase or 200 in L2$_1$ phase is observed for $T_{sub}$ = 600°C and 400°C, whereas no reflection is observed for $T_{sub}$ = RT. Note that, due to the absence of strong reflections from MgO in this experimental setting, we can exclude a possibility that the reflection is multiple scattering of MgO. A preliminary powder neutron diffraction experiment for bulk

Mn$_2$VAl at high temperatures has confirmed that 100 reflection in A2 phase at room temperature originates from antiferromagnetic (AFM) ordering [16]. Thus, along with the discussion in the last paragraph, we conclude that 1) Mn$_2$VAl film deposited at 600°C is L2$_1$ phase, 2) one deposited at 400°C is A2 phase with AFM ordering, and 3) one deposited at RT is A2 phase without AFM ordering. A summary of the neutron experiments is shown in Table 1.

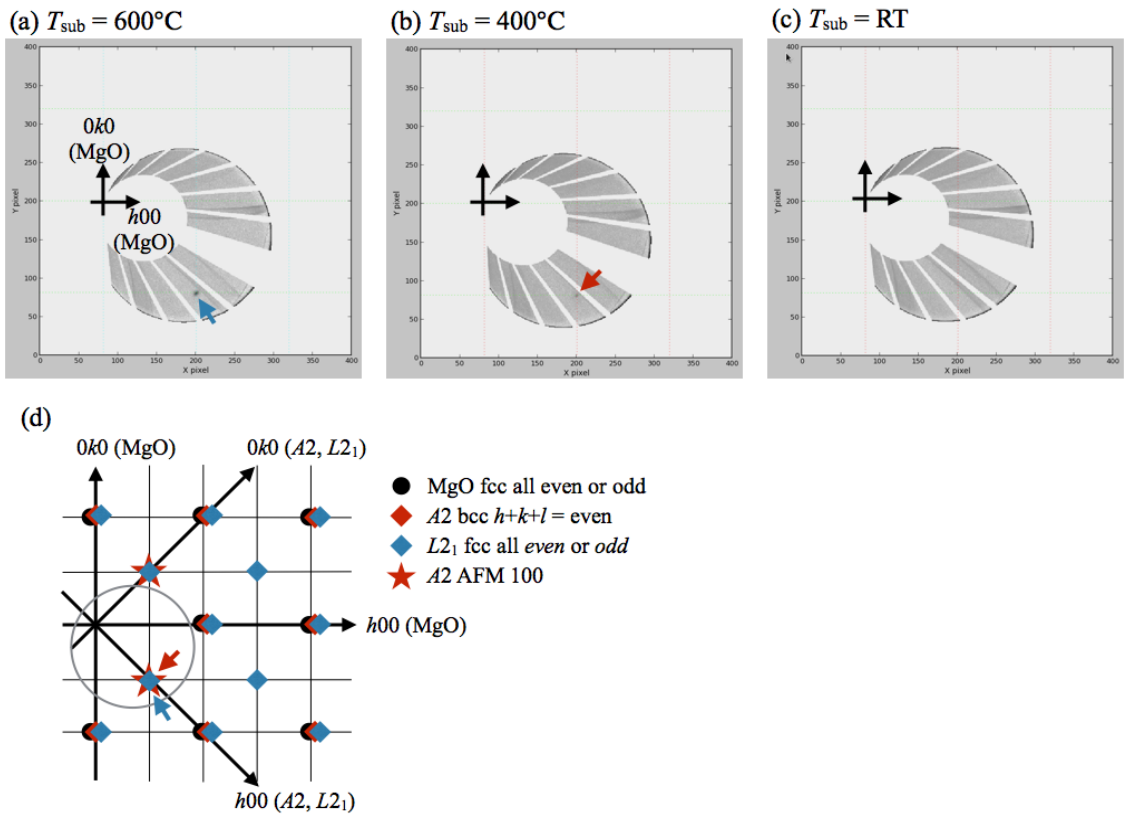

FIG. 4 Bragg reflections in the ($hk$0) planes of A2 and L2$_1$ phases. (a)-(c) Observed reflections for Mn$_2$VAl films with $T_{sub}$ = 600°C, 400°C, and RT. (d) Schematic illustration of the observed reflections indicated by blue and red arrows for L2$_1$ and AFM A2 phases, respectively.

TABLE 1. A summary of the neutron experiments on $Mn_2VAl$ films.

| Deposition temperature $T_{sub}$ | Observed reflections | | Chemical phase and magnetic structure |
|---|---|---|---|
| | $L2_1$ reflections in Fig. 3 | $L2_1$ 200 or A2 100 (AFM) in Fig. 4 | |
| RT | No | No | A2 para |
| 400°C | No | Yes | A2 AFM |
| 600°C | Yes | Yes | $L2_1$ |

## D. Exchange bias in $Mn_2VAl$/Fe bilayers

EB effects in $Mn_2VAl$/Fe bilayer samples were evaluated using A2-type $Mn_2VAl$ films with $T_{sub}$ = RT and 400°C. Both bilayer samples showed no EB shift at RT. *M-H* curves measured at 10 K are shown in Figure 5. Samples were cooled down to 10 K from 300 K under a magnetic field of 10 kOe before starting measurements. At 10 K, no EB shift appears for $T_{sub}$ = RT, whereas a shift of 120 Oe appears for $T_{sub}$ = 400°C. Although the blocking temperature is low and the shift is too small for device application at this moment, EB has been successfully demonstrated using the newly developed AFM $Mn_2VAl$ film.

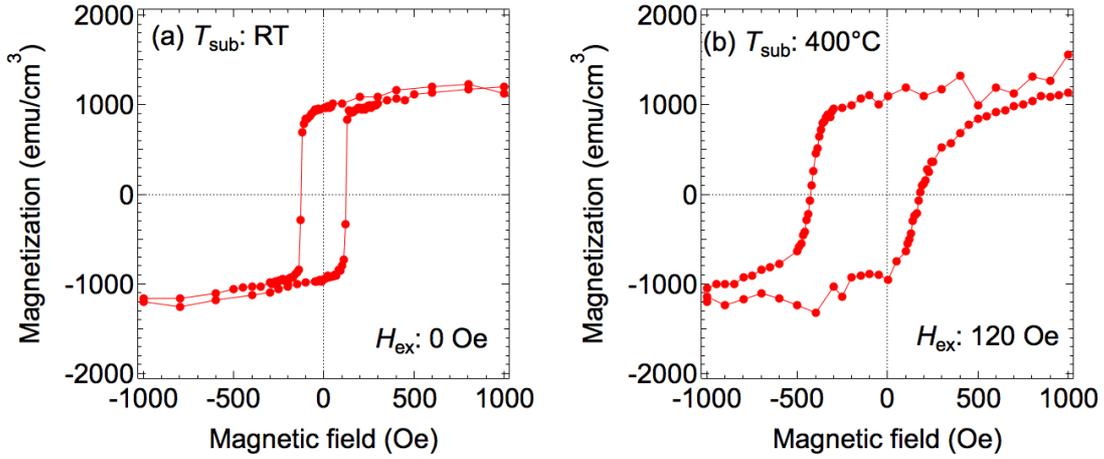

FIG. 5. *M-H* curves of A2-type $Mn_2VAl$/Fe bilayer samples measured at 10 K. The deposition temperatures, $T_{sub}$, for the $Mn_2VAl$ layer are (a) RT and (b) 400°C.

## IV. CONCLUSIONS

The structural and magnetic properties of epitaxially grown $Mn_2VAl$ films were investigated. The chemical phases of the films were successfully controlled by changing $T_{sub}$. The maximum saturation magnetization of 220 emu/cm³ was achieved in a film with $L2_1$ phase, which is larger than a reported value probably due to the improved chemical ordering with better stoichiometry. An A2-type $Mn_2VAl$ film deposited at 400°C showed magnetic superlattice reflection associated with the AFM phase in the neutron diffraction experiments at room temperature, and a $Mn_2VAl$/Fe bilayer sample with an AFM A2-type $Mn_2VAl$ layer deposited at 400°C showed an exchange bias shift of 120 Oe at 10 K.

## ACKNOWLEDGEMENTS

This work was partially supported by Heusler Alloy Replacement for Iridium (HARFIR) under the Strategic International Cooperative Research Program (SICORP) from Japan Science and Technology Agency (JST). A part of this work was done by a cooperative program (No. 16G0407) of the CRDAM-IMR, Tohoku University. Authors would like to thank Dr. R. Y. Umetsu for valuable discussion, and TT and TK would like to thank Mr. I. Narita for his technical support. TT acknowledges a support from the Graduate Program of Spintronics (GP-Spin) at Tohoku University. RK acknowledges a support from the Leading Graduates Schools Program, "Interdepartmental Doctoral Degree Program for Multi-dimensional Materials Science Leaders" by the Ministry of Education, Culture, Sports, Science and Technology. The neutron experiment at the Materials and Life Science Experimental Facility of the J-PARC was performed under the Elements Strategy Proposal (No. 2015E0003).